# Development of an Experimental Setup to Analyze Carbon/Epoxy Composite Subjected to Current Impulses

P. GHARGHABI, J. LEE, M. S. MAZZOLA and T. E. LACY

**ABSTRACT**

The use of composite materials is significantly growing in the aircraft industry because of their lightweight, high specific mechanical properties, such as strength and stiffness. These benefits are a potential improvement in aircraft design to save weight and reduce fuel consumption. A carbon fiber reinforced polymer (CFRP) composite is one of the types used for aircraft structural applications. Carbon fibers are mechanically stronger and lighter than aluminum. In a CFRP composite, carbon fiber acts as reinforcement and provides mechanical strength, and epoxy resin is the supporting matrix. Therefore, carbon fibers greatly improve composite's overall material properties. They can be fabricated in a desired stacking sequence, with various fiber orientations. From an electrical point of view, a composite can be designed so that it will conduct current asymmetrically in one direction as compared to another. A fiber orientation in a CFRP composite results in strong anisotropic material properties. This indicates that the composite exhibits different conductivity in each direction, along the fibers or transverse to the fibers.

In this paper, variation in electrical properties of CFRP caused by relatively low magnitude current impulses discharged through CFRP coupons are reported, and internal changes of the composite, due to current impulses, are studied. Based on electrical resistance measurements caused by these currents, property changes in CFRP composite coupons of two different carbon ply orientations are compared. Furthermore, it will be investigated if the changes caused by current impulses are mainly focused in the region of the current injection at the edges of composite coupons, or are they distributed uniformly through the bulk of the composite.

Pedram Gharghabi, Department of Electrical and Computer Engineering, Mississippi State University, MS 39762.
Juhyeong Lee, Department of Aerospace Engineering, Mississippi State University, Mississippi State, MS 39762.
Michael S. Mazzola, Department of Electrical and Computer Engineering, Mississippi State University, MS 39762.
Thomas E. Lacy, Jr., Department of Aerospace Engineering, Mississippi State University, Mississippi State, MS 39762.

# INTRODUCTION

Carbon fiber reinforced polymer (CFRP) composites have been widely used for many engineering applications. They exhibit advanced structural capabilities, such as low density, corrosion and fatigue resistance, high stiffness, and high strength. Another major benefit is better fatigue performance of CFRP compared to traditional metal (*i.e.*, Al, Steel). These properties make CFRP composites a suitable and attractive choice in aerospace industry and wind turbine manufacturing. Carbon fiber composites are gradually replacing aluminum structure in the design of aircraft, which is effective for weight saving and fuel consumption reduction. However, the emergence of composite material as the main structural component of aircraft introduces new considerations and problems that need careful study.

Conventional airframes made from metallic material such as, aluminum or copper, are not as vulnerable to lightning strikes. This arises from the fact that metallic materials are inherently conductive, and they keep the impulse current on the exterior surface of the aircraft and prevent it from penetrating through the skin of the aircraft where it could inflict serious damage. However, flow of current can damage the aircraft structure by burning and melting at the lightning strike attachment point.

Modern aircraft are increasingly constructed from composite materials such as, carbon fiber composites. When the aircraft is hit by a lightning strike, the composite surface suffers from direct and indirect effects. Puncture and burn through are types of direct effects of lightning strike that occur at the attachment point. Residual areas undergo indirect effects of lightning strikes that causes hidden failures.

Carbon fibers are relatively good conductors, while the polymer matrix acts primarily as a dielectric that increases the resistivity of the CFRP composites. There can be some damage to the aircraft structure after a lightning strike, and they can be categorized as direct effects and indirect effects. When the lightning strike hits the aircraft, a highly time varying impulse of current spreads out through the wings, fuselage and exterior skin of the aircraft, until it exits the aircraft. A typical lightning bolt carries an electric current of 30,000 Amperes [1]; lightning currents over 200 kA have also been recorded. As the current flows through the body of the aircraft, the current density and its magnitude decreases. Thus, areas away from the attachment point, which can be called residual areas, experience much lower current density.

Many papers have studied the current flow and composite response subjected to different profile of currents. In [2], voltage current characteristic of a composite versus time subjected to direct current (DC) and temperature rise caused by flow of DC current were investigated. It was shown that the conductivity of the composite is independent of the magnitude of applied DC voltage.

Conductivity of a composite is determined by carbon fiber volume fraction [3], and any change in the conductivity indicates an internal change in the composite, such as delamination or fiber fracture. Some reports [4, 5] proposed that conductivity of the composite can be used as a health monitoring criteria or self-sensing method. In [6] stress-strain testing was applied to the composite, and change of resistivity was measured. It was observed that by applying load on the composite, it would first increase linearly due to elongation of carbon fibers. Then, there would be a rapid increase in resistivity because of carbon fracture.

Several papers studied the direct effect of lighting strike on epoxy composite laminate by creating artificial lightning [7, 8, and 9]. Y. Hirano et al. [9] investigated the damage to the carbon fiber epoxy resins caused by a lightning strike and relationship between lightning characteristics and damage response. Gagne [10] proposed that smart materials such as polymer based composites are a suitable candidate to replace conventional lightning strike protection (LSP) technologies, because they are lightweight and mechanically reliable, and they meet current regulations and standards.

It is known that anisotropic materials have different electrical and thermal properties in each direction [11]. This fact makes it harder to study the response of these materials to electrical current flow and determine current flow distribution. Chippendale [12] demonstrated that comprehensive characterization of the material is very important to be able to accurately estimate the current flow in composites.

Hart et al. [13] studied the electrical resistance of composites when exposed to a high-intensity pulsed electric field. It was concluded that electrical resistance decreases with an increase in the current peak amplitude. It was also found that stacking sequence of the composite and its thickness have a great influence on its electrical resistance. In [14] delamination, cracks, and joints are good examples of complex feature in composite structure. Not as much research has been carried out to study the effect of lightning strikes on composite materials beyond the attachment point. Damage that can change the structural integrity of CFRP composite can be caused by highly time-varying current beyond the attachment point that is not visible.

The focus of this study is not the direct effect of lightning on a composite at the attachment point, but rather the areas away from the attachment point that experience lower magnitude of the impulse current. Bearing in mind that the dimension of the specimens used in this and many other studies are much smaller than the dimension of an aircraft, the magnitude of the current that the coupon experiences needs to be scaled down proportionally.

In this paper, current impulses were applied to composite coupons using a current impulse generator. After which, changes in resistance were measured. The objective of this paper is to observe the change of conductivity due to low magnitude current impulses simulating the lightning current in an aircraft structure after spreading away from the attachment point.

Carbon fiber composites are highly anisotropic materials that have different characteristics in each direction and are fabricated in various stacking sequences with fibers origination in different directions. Flow of current through the body of a composite specimen may change its electrical and mechanical parameters. Electrical resistance of the specimen is a basic parameter that can be caused by an internal change in the body of the specimen. Therefore, to better understand the behavior of the specimens when they are subjected to impulse current, the change of resistance against magnitude of current impulses is studied.

**EXPERIMENTAL METHODS**

To perform current impulse tests on composite coupons a system was designed which is capable of producing current impulses up to 300 A peak amplitude in the range of microseconds. The schematic of the system is shown in Figure 1.

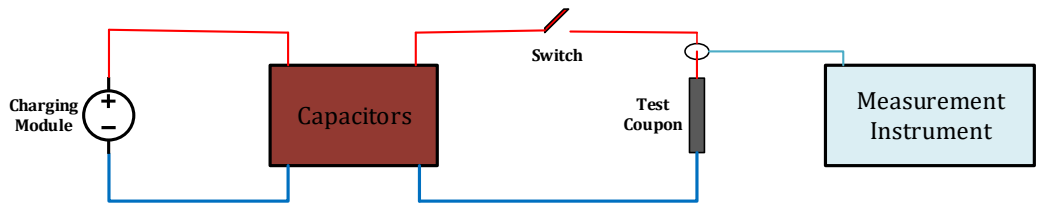

Figure 1. Schematic of current impulse test setup

The test setup consists of a charging module, capacitors for energy storage, a test coupon, electrical contacts, and a measurement instrument. The charging module is a DC source that can provide voltages up to 1 kV, and it feeds the capacitors through a gate driver up to the target voltage. The capacitor bank consists of four 100 µF capacitors, and they are connected in parallel to ensure that maximum current can be derived from this setup. An enabling component of the impulse generator is a 1200 V SiC JFET multi-chip transistor module. This module can safely switch impulse currents greater than 1 kA at a di/dt in excess of the wave shape requirements for this study. High di/dt switching of the module is enabled by a proprietary gate drive circuit designed specifically for SiC JFET modules.

The current impulse was measured using a clamp current measuring device, which is a current transformer with wire coil as one winding and conductor as the other winding. The output of the clamp is connected to a 1 GHz digital oscilloscope, which captures and records the current wave shape. By triggering the gate driver, the energy stored in the capacitor bank is discharged to the load. The higher the charging voltage, the larger the output current will be. Therefore, by controlling the input voltage the discharged current impulse can be changed to the target value. This fact is based on the valid assumption that all the electrical parameters of the circuit remain constant. The impulse current wave form recorded at charging voltage of 300 V applied on the unidirectional coupon sample 1 is shown in Figure 2. Different current impulse parameters can be measured from the waveform. The peak amplitude of the current impulse is measured to be 284 A, and rise time of the waveform, which is defined from zero to peak amplitude, is measured to be 4.68 µs. The tail time, from the peak amplitude to half of the peak value on the tail of the waveform, is 17.6 µs.

The test was conducted at different voltage levels. Starting from the lowest charging voltage of 50 V, the test was repeated 3 times at each level. Then, the voltage was increased in steps of 50 V until the current limit, below 300 A, was reached. At every shot, current wave shape was recorded, and the resistance of the coupon was measured.

Providing a proper electrical contact between the electrode and the coupons is very important to minimize the heat generated at the interface of electrode and composite coupons. When the current impulse is injected into the composite coupon, the flow of current through the contact surface generates heat that heats the contact up. High resistance at the contact, disturbs the consistency of the measurements and influences the results. It might also lead to burning the epoxy matrix at the interface. To reduce the contact resistance, a thin layer of epoxy on the top the coupon was removed using sand paper to expose carbon fibers and ensure smooth contact surface. A copper strip was placed between composite coupon and the contact. It was firmly clamped to ensure a strong and low resistance electrical contact.

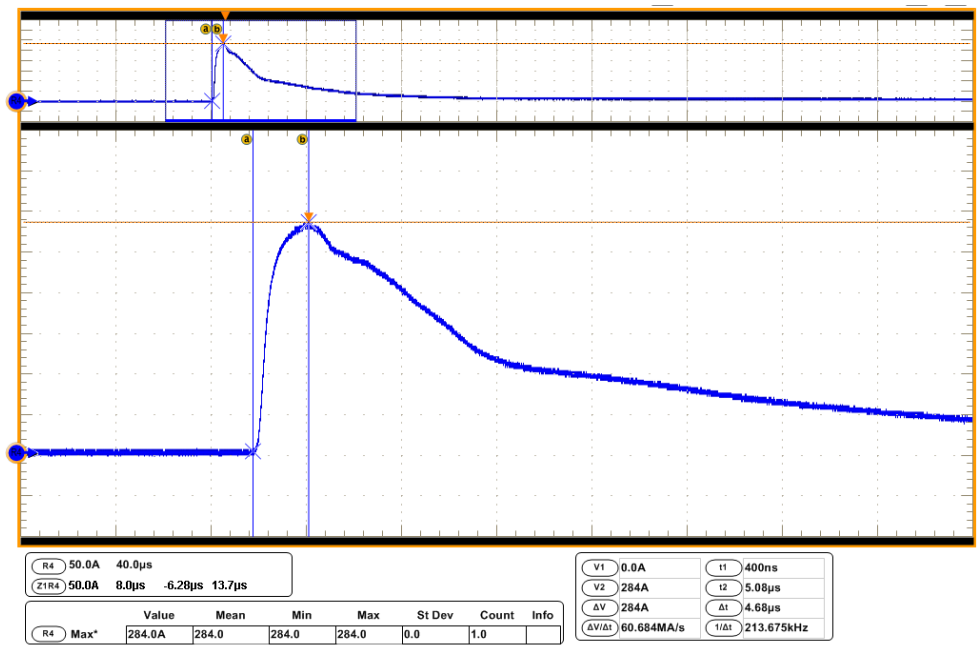

Figure 2. Electric current vs time, unidirectional coupon, sample 1, peak amplitude 284 A front time 4.68 μs, tail time 17.6 μs

After completion of the current tests, the coupons were cut crosswise into three sections and the resistance of each section was measured and recorded. The electrical resistance of each section were compared to localize where the apparent internal change occurred in the original coupon. The two sections that were in contact with electrodes represent the attachment point current is injected and current removed. The center section can be considered isolated from the region of the current injection and extraction, and it is of interest to see if the overall change in coupon resistance is primarily located in the end sections, or in the middle sections, or equally distributed in all sections. The hypothesis is that the two sections that are in contact with the electrode, accumulate more change due to the low-current impulses. Our interest in testing this hypothesis is to take an initial step toward ruling in or ruling out possible mechanisms for the resistance change. The reasoning is that if the resistance change is found to be located preferentially in the two end sections, this is evidence that current is crossing ply/resin layer boundaries near the contact interface and producing changes to the polymer bonding with the carbon fiber matrix before diverging laterally to flow along fibers through the center section. This is the underlying mechanism proposed in this paper to explain the permanent, reproducible, and monotonic reduction in resistance caused by very low current densities as compared to that in the neighborhood of a lightning strike attachment point.

**BASICS CIRCUIT ANALYSIS**

The impulse current generator can be represented by a series RLC circuit to study its response and output specifications. The equivalent electrical circuit is shown in Figure 3 and all three components are in series. C represents the equivalent capacitance in the circuit, since four capacitors are connected in parallel and the equivalent capacitance would be 400 µF. L represents the self inductance in the circuit caused by loop of wires or straight wires and connections, and $V_0$ is the DC charging voltage supplied by the charging unit. Using Kirchhoff's voltage law, the governing differential equation can be written as

$$\frac{1}{C}\int i\, dt + L\frac{di}{dt} + Ri = V_0$$
$$\frac{d^2i}{dt^2} + \frac{R}{L}\frac{di}{dt} + \frac{1}{LC}i = 0,\ i(0^+) = 0,\ \frac{di(0^+)}{dt} = V_0/L \tag{1}$$

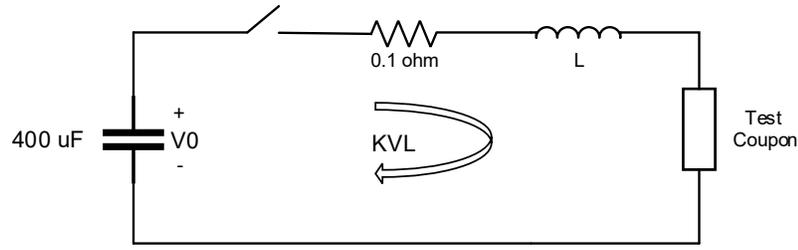

Figure 3: Series RLC circuit representing the current generator

The characteristic equation can be expressed in general standard form of second order linear systems

$$\frac{d^2i}{dt^2} + 2\alpha\frac{di}{dt} + \omega_0^2 i = 0 \ \rightarrow\ \alpha = \frac{R}{2L},\ \omega_0 = \frac{1}{\sqrt{LC}},\ \varsigma = \frac{\alpha}{\omega_0} = \frac{R}{2}\sqrt{\frac{C}{L}} \tag{2}$$

Where α and ω₀ are attenuation and undamped natural frequency, both in angular frequency. Larger attenuation forces the wave shape closer and decay quicker. ζ is damping ration of the circuit. Depending on the value of ζ the output can have different patterns.

The first return stroke of lightning current is approximated by a double exponential wave shape. To have the overdamped case (i.e., double exponential waveform), ζ should be greater than 1. In this case, the two poles are real and unequal on the left half side of S plane. The solution of the equation may be found as

$$s_{1,2} = \left(\varsigma \pm \sqrt{\varsigma^2 - 1}\right)\omega_0,\ i(t) = A\left(e^{-s_1 t} - e^{-s_2 t}\right) \tag{3}$$

**Specimen Preparation**

The composites were fabricated using 8 plies of carbon/epoxy prepreg (TCR prepreg carbon unidirectional tape, SGL C30 T050 EPY 50K/UF3325 carbon fiber/epoxy resin system). Two different material configurations were compared to characterize the effect of electrical conduction path, which is associated with a stacking sequence, on electrical resistances of the composites: (1) a [0] unidirectional composite and (2) a $[0/90]_{4n}$ cross-play composites. Each lamina has in-plane dimensions of 20 cm × 25 cm (width × length) with a nominal thickness of 0.685 mm. These composites were cut into four 5 cm × 25 cm test coupons using a water-cooled diamond coated circular saw. In order to compare a change in electrical resistance depending on locations, each of test coupons were then cut into three 5 cm × 8.3 cm test coupons. In this study, three specimens were subjected to current impulses and one specimen was given for reference.

**RESULTS AND DISCUSSIONS**

One of the main parameters of lightning strike that inflicts damage to any object is the magnitude of the current stroke. The heat generated due to flow of current at the attachment point and areas far from the attachment point is mainly dependent on the magnitude of the current impulse. In this study, the magnitude of discharged current is the parameter that plays an important role in resistance change in the specimen.

The change of electrical resistance of the unidirectional and cross ply coupons are shown in Figure 4 and Figure 5. The trend of the change of electrical resistance against the cumulative number of current impulses shows that successive current impulses changes the resistance of the coupons. Similar phenomenon was also reported in [13], and it was concluded that electrical resistance decreases with increase in the applied current magnitude. The decrease of the resistance is indicative of internal change in the carbon composite which is irreversible.

To be able to compare measurements of two unidirectional and cross ply coupons, the values and measurements should be normalized or adjusted to a common scale. The average resistances of three coupons at each level of charging voltage is divided by the average resistance before any impulse application. Therefore, all readings are between 0 and 1, which shows normalized resistance change. The diagram of normalized resistance change against magnitude of current impulse is presented in Figure 6.

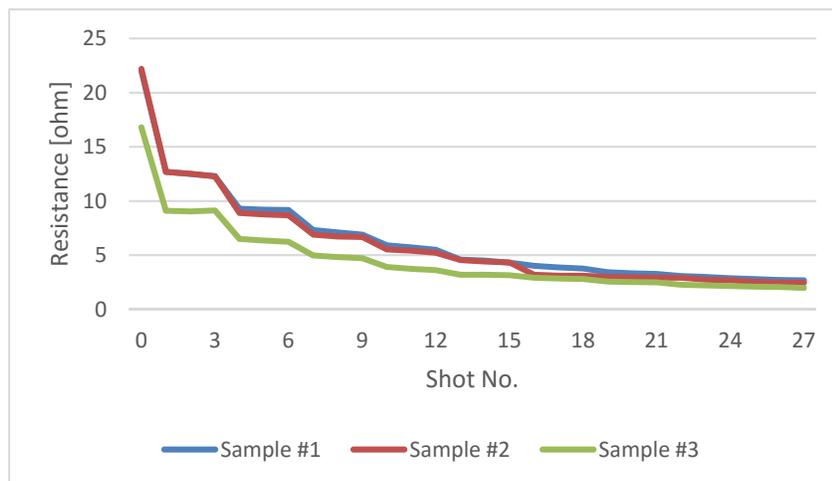

Figure 4. Resistance measurements of current impulse test,
Cross ply coupons, sample No. 1, 2 and 3

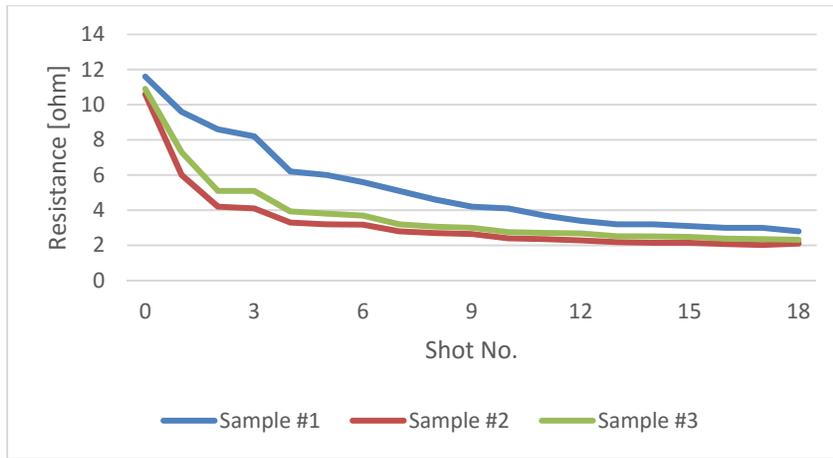

Figure 5: Resistance measurements of current impulse test, Unidirectional coupons, sample No. 1, 2 and 3

Because the unidirectional coupons have lower resistance, they reach higher discharge currents at lower voltages. Thus, cross ply coupons need higher voltage to reach current limit. The resistance of both samples decreases as the magnitude of current impulse increases. They reach almost half of initial value of resistance at roughly around 10 A of discharged current, which is a relatively low current compared to actual lightning impulse. To explain the relevance of this testing in that context, we consider a coupon is a small patch of a much larger portion of the aircraft structure that is carrying lightning impulse current in a zone away from the attachment point; which makes the current levels used in this study relevant.

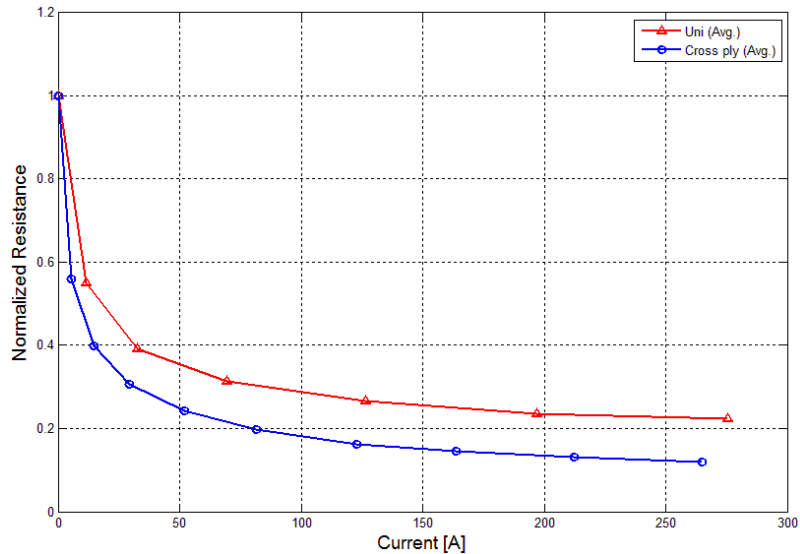

Fig. 6: Normalized electrical resistance vs current discharge magnitude, Cross-ply and unidirectional coupons

Before any application of current impulses to the specimens, the average value of electrical resistance of three cross ply laminates is 20.26 Ω, while that of unidirectional laminates is 11.03 Ω. As demonstrated in Figure 5, the resistance decreases considerably during the first few current impulse applications. After this significant drop, the resistance change lessens, and tends to be linear at higher current impulses. This trend, at higher current magnitudes, might imply that the path for the current flow is formed inside the coupons. Another result that can be deduced from Figure 5 is that cross ply coupons experience more resistance change compared to unidirectional coupons. In a cross ply coupon, discontinuity of different electrical and mechanical properties at the interface between carbon fiber and epoxy matrix is more pronounced. Therefore, the effects of current crossing laminate boundaries are more accentuated, and the effect of the weak zones on the resistance change inside coupons is greater in cross ply coupons as compared to unidirectional coupons.

**Resistance Measurement after Current Application**

The principal experimental step of this study was to begin the process of determining if the overall change in electrical resistance of each coupon is localized in regions of the coupon, such as the edges near the contact interfaces or away from them toward the middle (or "bulk") of the coupon. Each specimen is segmented in to three section, 1 and 3 (edge sections), and 2 (middle section). The measurements are presented in Table I. It shows the resistances of the coupons before any current application and after current application, and in the following three columns it shows resistances after sectioning the coupons in to three equal parts.

Table I shows that after application of current impulses, in 5 coupons out of 6 the middle section has higher resistance compared to edge sections. Therefore, it is safe to say that sections 1 and 3, that have the electrical contact where the current impulse is directly injected normal to the surface of the coupon, experienced more damage as compared to middle sections. The coupons that are not tested and kept as a reference, show almost equal resistance on the three sections. In theory, summation of resistances of 3 sections should be equal to the resistance of a coupon after application of current impulses before sectioning.

TABLE I. RESISTANCES MEASUREMENTS OF TEST SAMPLES AFTER SECTIONING

| Sample | Resistance before current impulses [ohm] | Resistance after current impulses [ohm] | Resistance of section 1 [ohm] | Resistance of section 2 [ohm] | Resistance of section 3 [ohm] |
|---|---|---|---|---|---|
| Cross ply not tested | 23.2 | - | 9.4 | 9.4 | 8.8 |
| Cross ply #1 | 22.0 | 2.68 | 1.42 | 2.8 | 1.36 |
| Cross ply #2 | 22.20 | 2.46 | 1.22 | 3.05 | 1.12 |
| Cross ply #3 | 16.80 | 1.98 | 1.42 | 2.75 | 1.38 |
| Uni not tested | 9.7 | - | 2.5 | 2.5 | 3 |
| Uni #1 | 11.60 | 3.0 | 1.6 | 1.7 | 1.4 |
| Uni #2 | 10.60 | 2.10 | 1.32 | 1.65 | 1.78 |
| Uni #3 | 10.90 | 2.32 | 1.12 | 1.84 | 1.71 |

**CONCLUSIONS**

The objective of the present study is to investigate the electrical behavior of the composite coupons when they are subjected to low magnitude current impulses. A custom-built setup was developed to generate the current impulses, measure and record the current wave shapes, and measure the electrical resistances of samples. Two composite samples were made: a unidirectional sample and a cross ply sample, and cut into equal coupons. A series of current impulses were applied to the coupons and the response of the composite samples were investigated. Change in the electrical resistance of the samples is evidence of internal change in the material properties of the composite coupons, an assertion that will be investigated by mechanical testing of effected coupons in on-going work to be reported in the near future. It was found that flow of low-magnitude current impulses have an unexpectedly significant effect on the material property of the composite samples; as it was observed that successive current application irreversibly reduces the resistance of the coupons. The incremental change of resistance is considerable at the beginning of current applications and slows as current impulses accumulate suggesting a saturation of the irreversibly changed internal properties.

Cross ply coupons are more sensitive to current impulses, and they exhibit more change in electrical resistance. This might be attributed to the orthogonal discontinuity of conducting versus insulating material forming the laminations. When these boundaries are positioned with respect to the natural direction of current flow the carbon fiber and epoxy matrix, may form dielectric weak zones inside composite material due to changes in polymer/fiber bonding and/or interfacial dielectric properties, the cumulative result seeming to be a reduced barrier to current flow at these boundaries than those that exist in the virgin coupons. As an initial test of this hypothesis, after completion of current impulse applications the coupons were cut into three equal sections. The measurement of electrical resistances confirms that the observed change of resistance preferentially occurs in the sections that are in contact with the electrode, although Table I indicates this effect is more pronounced with the cross ply coupon than the unidirectional coupon. In the unidirectional coupon the divergence of the current from normal at the surface of the contact interface to parallel with the ply direction in the center of the coupon is less pronounced and may reduce the localization of the effect. Our working hypothesis is that this is further evidence that current flowing across resin/ply laminate interfaces is causing changes in the dielectric strength of these interfaces. This explanation is consistent with the additional observations of a greater absolute and relative change in resistance for the cross ply coupon as compared to the unipolar ply coupon.

However, much more work needs to be done to understand this surprising effect that has implications on the possible effects of lightning impulse currents on carbon composite materials in areas of the aircraft structure far removed from the lightning attachment point.